\documentclass[sigconf]{acmart}

\AtBeginDocument{%
  }

\usepackage{fontawesome}
\newcommand{\Video}{\href{https://www.youtube.com/watch?v=e0xtntdpcmg&ab_channel=Golaem}{\faYoutubePlay}}
\setcopyright{rightsretained}
\copyrightyear{2024}
\acmYear{2024}
\acmConference{SIGGRAPH Talks '24}{July 27 - August 01, 2024}{Denver, CO, USA}\acmBooktitle{Special Interest Group on Computer Graphics and Interactive Techniques Conference Talks (SIGGRAPH Talks '24), July 27 - August 01, 2024}\acmDOI{10.1145/3641233.3664312}
\acmISBN{979-8-4007-0515-1/24/07}

\begin{document}

\title{Audio2Rig: Artist-oriented deep learning tool for facial animation}

\author{Bastien Arcelin}

\orcid{1234-5678-9012}
\affiliation{%
  \institution{Golaem}
  \city{Rennes}
  \country{France}
}
\email{bastien.arcelin@golaem.com}

\author{Nicolas Chaverou}
\orcid{0009-0003-8659-0880}
\affiliation{%
  \institution{Golaem}
  \city{Nouméa}
  \country{New Caledonia}}
\email{nicolas.chaverou@golaem.com}


\begin{abstract}
  Creating realistic or stylized facial and lip sync animation is a tedious task. It requires lot of time and skills to sync the lips with audio and convey the right emotion to the character's face. To allow animators to spend more time on the artistic and creative part of the animation, we present \textit{Audio2Rig}: a new deep learning based tool leveraging previously animated sequences of a show, to generate facial and lip sync rig animation from an audio file. Based in \textit{Maya}, it learns from any production rig without any adjustment and generates high quality and stylized animations which mimic the style of the show. 
  \textit{Audio2Rig} fits in the animator workflow: since it generates keys on the rig controllers, the animation can be easily retaken. 
  The method is based on 3 neural network modules which can learn an arbitrary number of controllers. Hence, different configurations can be created for specific parts of the face (such as the tongue, lips or eyes). 
  With \textit{Audio2Rig}, animators can also pick different emotions and adjust their intensities to experiment or customize the output, and have high level controls on the keyframes setting. Our method shows excellent results, generating fine animation details, respecting the show style. Finally, as the training relies on the studio data and is done internally, it ensures data privacy and prevents from copyright infringement. Video examples are available \Video.
\end{abstract}



\begin{teaserfigure}
  \includegraphics[clip, width = \textwidth]{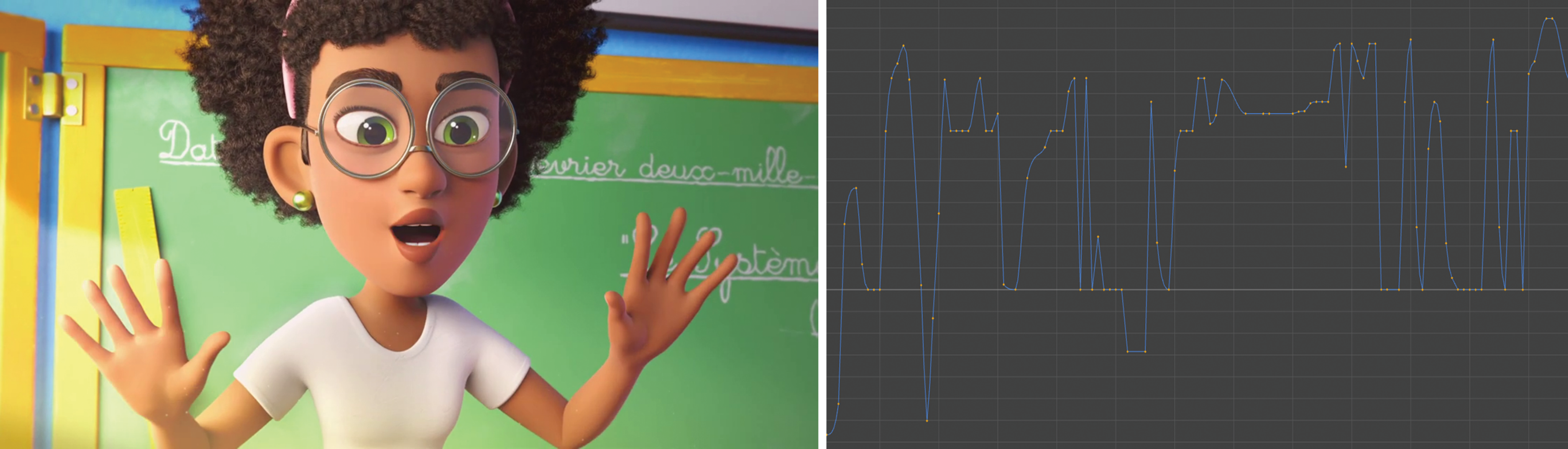}
  \Description{Example of facial and lip sync animation (left) and view of keys set on a facial rig controller (right). © Stim Studio}
  \label{fig:results}
\end{teaserfigure}


\maketitle

\section{Related works}
Some off-the-shelf tools are able to produce facial animation from audio files: \textit{Audio2Face} \cite{Karras17} from \textit{Nvidia} uses audio to animate vertices of a mesh, and \textit{Polywink} proposes semi-automatic face rigging and phoneme based animation from audio. Yet, neither of them are able to operate on production rigs without adjustment nor take the show style into account. More importantly, the results can not be easily retaken due to the lack of high level controls. More recently, \cite{blizzard} proposed to extract phonemes from audio using a pre-trained Neural Network (NN) and map those to predefined controller postures. However, this method requires the creation of a facial shape for each phoneme and the resulting animation needs to be retaken to adjust for emotion or language speed. It also does not take the show style into account. 

In this paper, we propose a tool which learns from any facial rig and from the animator's work. Therefore, it  generates high quality animations in the style of the show. \textit{Audio2Rig} provides high level controls to the animators and produces animations that can be easily retaken since it sets keys on the controllers.

\section{Audio2Rig}


\subsection{Data Extraction and Processing}
\begin{figure}
  \centering
  \includegraphics[trim = {0.5cm 9.5cm 0.5cm 0cm}, clip, width = 0.9\columnwidth]{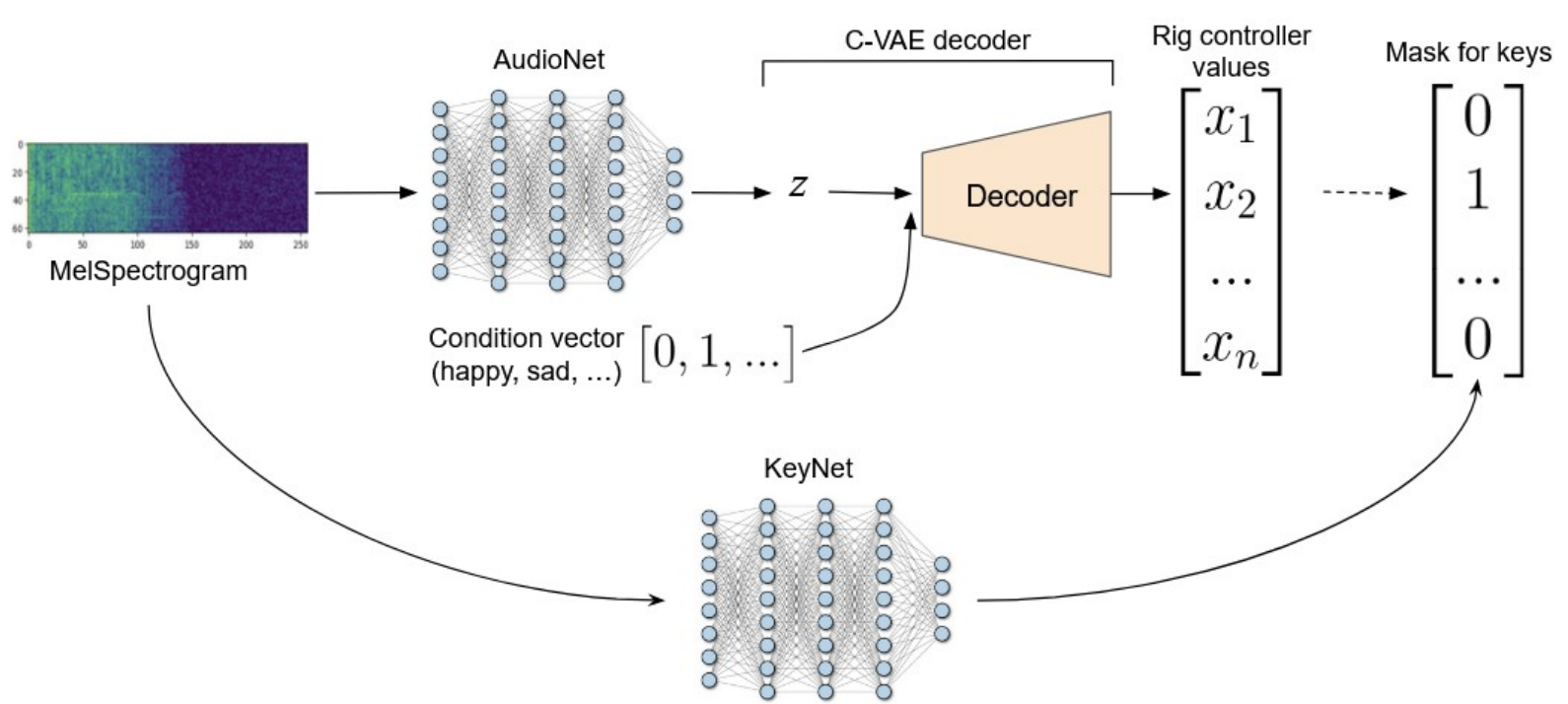}
  \caption{Audio2Rig inference workflow}
  \Description{Audio2Rig inference workflow}
  \label{fig:net_arch}
\end{figure}
To test our tool, we relied on 9471 frames of animation of a character provided by Stim Studio and animated by Blue Spirit Studio. The data is composed of 6 different animated sequences, each with a different emotion. 
We extracted the controllers values at each frame to use them as inputs and / or outputs of the networks during training. Since this input can be pretty sparse, all controllers not keyframed are removed before training, preventing the NN to converge to a zero mean. Before feeding the audio to the NN, it is to transformed into a MelSpectrogram, a 2D input well suited to Convolutional Neural Networks (CNN) which facilitates features extraction. 
The transformed audio is split into fixed size time intervals centered on the target animation frame.

\subsection{Neural Network Architecture}
A set of 3 NN are used in this method. First, a Conditional Variational AutoEncoder (C-VAE, \cite{cvae}) is trained. It is a dense NN which learns to reproduce the facial rig controllers values. It maps these values into a latent space \textit{z} and generates them from this latent vector, taking into account a \textit{N} dimensional condition vector \textit{c} (here, $\textit{N}=6$). This vector represents the different possible emotions. Then, a CNN (AudioNet) learns to process the audio signal. It is composed of several convolutional layers, one Gated Recurrent Unit layer \cite{gru}, and dense layers. It is trained to predict the values of the C-VAE latent space, \textit{z}. The trained C-VAE decoder is used to infer the controllers values taking into account a condition vector, \textit{c}. Finally, another CNN (KeyNet) is trained to learn from audio when to place keys for each controller. It also learns the values of the derivatives before and after each key. This network aims at reproducing the timing at which animators set keys, generating animations closer from the studio style. 

\subsection{Training Procedure}
\textit{Audio2Rig} is trained and deployed on a single character. It can be deployed on multiple characters so long as they have the same facial controllers and the user accepts stylistically diluted, less character-specific results. The training of C-VAE and AudioNet must occur sequentially. KeyNet can be trained in parallel.
\textit{Audio2Rig} allows flexible configurations: various sets of networks (C-VAE / AudioNet / KeyNet)
can be used for different parts of the facial rig. We trained 9 networks sequentially for 3 configurations: mouth, tongue and face upper part. It took 3h on a \textit{Nvidia GeForce RTX 4060} GPU. 

\subsection{Inference Workflow}
As shown on Fig.\ref{fig:net_arch}, AudioNet predicts a vector \textit{z} that is fed into the C-VAE decoder. This decoder predicts the controller values based on the condition vector, \textit{c}, set by the user using emotion settings (see \ref{section:emotion}). Then, KeyNet takes the audio as input and generates a mask that indicates when to apply a key on a controller, as well as the corresponding tangents values.

\subsection{High Level Settings}
\label{section:emotion}

\begin{figure}
  \centering
  \includegraphics[trim = {0.cm 14cm 0.cm 1cm}, clip, width = \columnwidth]{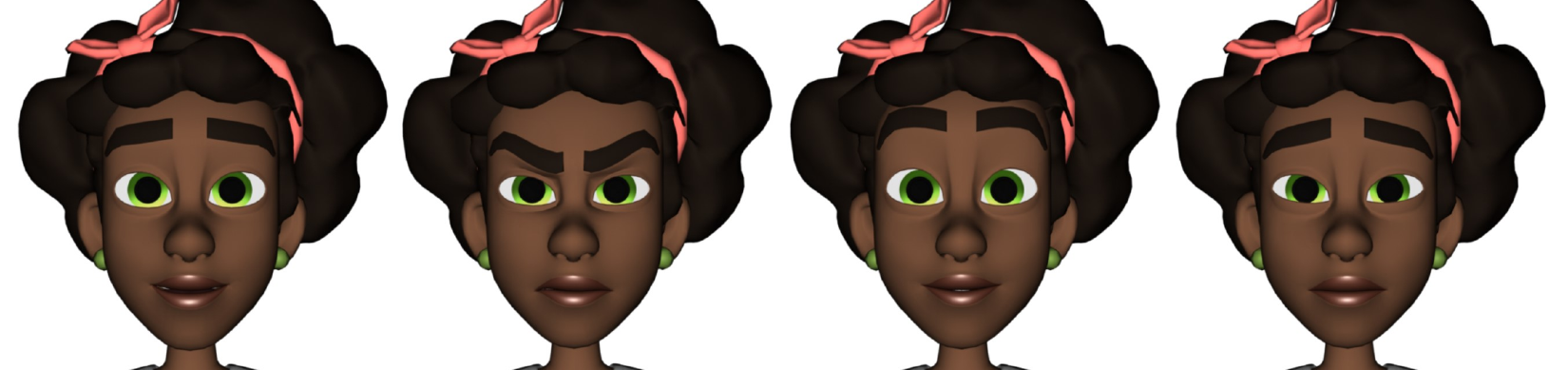}
  \caption{\centering{Example of applied emotions.\linebreak
  From left to right: happy, angry, surprised and sad.}}
  \Description{Example of applied emotions}
  \label{fig:emotions}
\end{figure}

\textit{Audio2Rig} provides different controls to the animators. First, they can decide which emotions to apply to the character using weights (Fig.\ref{fig:emotions}). Emotion weight varies between 0 and 1 and it is possible to mix emotions using several non-zero weight values. The rig is animated live during this step to provide a preview of the result and allow the artist to fine tune the settings for a more accurate animation. The weights values are used as the condition vector of the C-VAE at inference. They can also fine-tune how to set the keys. It is possible to filter predicted controller values or values of the animation curves tangents. These filters can be applied at different animation rates. As temporal coherence is harder to achieve on the face upper part, a Gaussian filter can be applied on the corresponding controllers, similarly to \cite{Karras17}.

\section{Results and Discussion}
We conducted experiments with 3 configurations of C-VAE / AudioNet / KeyNet sets
: the mouth, the tongue and the face upper part. \textit{Audio2Rig} generated accurate animations for the mouth and tongue. Regarding the face upper part, the emotion can clearly be identified. Yet, as it is more impacted by stochastic behavior (blinking in particular), the animation of the eyes or eyebrows remains less satisfactory. To improve temporal coherence on this part, we added a \textit{Smooth Upper Part} option in the animation settings.

In future work, we plan to support more emotions and the ability to apply keys on the emotion settings. It would allow animators to change the character's mood during a shot. We are also working on improving the face upper part animation accuracy. In parallel, we continue testing this method on more assets from various studios.

In conclusion, \textit{Audio2Rig} can be used to generate realistic and stylized facial and lip sync animation on any \textit{Maya} rig. Results can be retaken easily and it fits perfectly in the studio animation workflow. Since the tool learns from data that is produced within the studio, the accuracy of the networks will increase every time new data is produced for the show. As a side note, our tool is operating within \textit{Maya} but can be extended to other DCCs.

\bibliographystyle{ACM-Reference-Format}
\bibliography{talk_AudioToRig}




\end{document}